\documentclass[12pt]{iopart}

\newtheorem{lemma}{Lemma}
\newtheorem{proposition}{Proposition}
\newtheorem{theorem}{Theorem}

\newcommand{\dif}{\mbox{\rm d}}
\newcommand{\ci}{\mathop{\textrm{i}}}

\begin{document}

\title[Characterization of the Kerr metric] {An intrinsic
characterization of the Kerr metric}

\author{Joan Josep Ferrando$^1$\
%\footnote[3]{To
%whom correspondence should be addressed (joan.ferrando@uv.es)}
and Juan Antonio S\'aez$^2$}

\address{$^1$\ Departament d'Astronomia i Astrof\'{\i}sica, Universitat
de Val\`encia, E-46100 Burjassot, Val\`encia, Spain.}

\address{$^2$\ Departament de Matem\`atiques per a l'Economia i l'Empresa, Universitat de
Val\`encia, E-46071 Val\`encia, Spain}

\ead{joan.ferrando@uv.es; juan.a.saez@uv.es}

\begin{abstract}
We give the necessary and sufficient (local) conditions for a metric
tensor to be the Kerr solution. These conditions exclusively involve
explicit concomitants of the Riemann tensor.
\end{abstract}


%\submitto{\CQG}

%\maketitle

\section{Introduction}

The Kerr solution plays an essential role in relativistic
astrophysics to model the exterior gravitational field of a rotating
mass. Its prominence among stationary vacuum solutions comes from
the black hole uniqueness theorem which states that, under rather
general conditions, the Kerr space-time is the only asymptotically
flat, stationary, vacuum black hole.

Nowadays, the study of dynamical black hole space-times is based on
numerical simulations that usually make use of the 3+1 formalism.
Thus, it is important to understand the properties of the Kerr
metric on a generic space-like hypersurface. In a recent work
\cite{GP-VK-1}, the Cauchy initial data have been analyzed on a
generic slice of the Kerr geometry and a characterization of these
data is offered. This study is based on the space-time
characterization of the Kerr solution given by M. Mars \cite{mars1}
\cite{mars2}.

The Kerr initial data characterization given in \cite{GP-VK-1} fails
to be completely algorithmic as claimed by the authors. By contrast,
in a previous paper \cite{GP-VK-2} the same authors obtained a fully
algorithmic characterization of the Schwarzschild initial data. The
main reason for these differences is that the study of the initial
data made in \cite{GP-VK-2} is based on a local {\em intrinsic} and
{\em explicit} space-time characterization of the Schwarzschild
solution that we presented ten years ago \cite{fsS}. However, the
characterization of the Kerr geometry by Mars \cite{mars1}
\cite{mars2} used in \cite{GP-VK-1} does not possess such good
qualities. Nevertheless, these two works study the line suggested by
W. Simon \cite{simon} in depth, and have also been useful to improve
the black hole uniqueness theorem \cite{Ionescu}.

In this article we present a local {\em intrinsic} space-time
characterization of the Kerr solution that exclusively uses {\em
explicit} concomitants of the Weyl tensor. Our labeling of the Kerr
geometry is algorithmic and must allow an algorithmic
characterization of the Kerr initial data. But the relevance of this
intrinsic characterization goes further than the above mentioned
application.

Since the beginning of Riemannian geometry the invariant
characterization of metrics is a subject that has been tackled from
several points of view. Thus we have the historic theorems that
characterize locally flat Riemann spaces \cite{Riemann}, Riemann
spaces with a maximal group of isometries \cite{Bianchi}, and
locally conformally flat Riemann spaces \cite{Cotton} \cite{Weyl}
\cite{Schouten}. It is worth remarking that, in all these historic
results, the conditions involve explicit concomitants of the
curvature tensor (Riemann, Weyl and Cotton tensors). Then, they are
certainly {\em intrinsic} (depend solely on the metric tensor $g$)
and, besides, the {\em explicit} expression of these concomitants in
terms of $g$ is known.

Cartan \cite{cartan} showed that a Riemannian geometry may be
characterized in terms of the Riemann tensor and its covariant
derivatives. Brans \cite{brans} introduced the Cartan invariant
scheme in general relativity, and after Karlhede's work
\cite{karlhede}, this approach became more helpful within the
relativistic framework. The Cartan-Karlhede method to study the
equivalence of two metric tensors is based on working in an
orthonormal (or a null) frame, fixed by the underlying geometry of
the Riemann tensor.

Nevertheless, the historic theorems quoted above show that the
determination of a Riemannian canonical frame is not necessary in
labeling specific families of space-times. We find a similar
situation in the characterization of the Stephani Universes
(conformally flat perfect fluid solutions) or in the different ways
that the Friedmann-Lema\^itre-Robertson-Walker Universes can be
intrinsically labeled (conformally flat barotropic perfect fluid
solutions; perfect fluid solutions with vorticity-free and
shear-free geodesic velocity).

A suitable procedure is to analyze every particular case in order to
understand the minimal set of elements of the curvature tensor that
are necessary to label these geometries, an approach adapted to each
particular geometry we want to characterize. The examples quoted
above show that the characterization conditions usually involve
tensorial concomitants whereas the Cartan-Karlhede scheme only uses
scalar concomitants. Whatever the method, the covariant obtention of
the underlying geometry of the Weyl and Ricci tensors is a necessary
tool to characterize space-times intrinsically.

The algebraic classification of the space-time symmetric tensors was
given by Churchill \cite{churchill}, and since the sixties a lot of
work has been devoted to studying the Ricci tensor from an algebraic
point of view (see, for example, the Pleba\'nski paper
\cite{plebanski}). Nevertheless, the general covariant method to
determine the Ricci eigenvectors and their causal character was
presented by J.A. Morales in his Ph. D. Thesis (1988) and published
in \cite{bcm}.

The pioneering papers by Petrov \cite{petrov} and Bel \cite{bel}
studied the algebraic classification of the Weyl tensor. Since then
several algorithms have been proposed to determine the Petrov-Bel
type. The general covariant expressions that give the underlying
geometry of the Weyl tensor for every Petrov-Bel type were obtained
by J.A. S\'aez in his Ph. D. Thesis (2001) and published in
\cite{fms}.

The Kerr metric is a vacuum solution and, consequently, the Ricci
tensor vanishes, $Ric(g)=0$. On the other hand, the Weyl tensor $W =
Weyl(g)$ is Petrov-Bel type D. The first one is an intrinsic and
explicit condition in the metric tensor. The second one is intrinsic
and we must write it in an explicit form. Thus, in the first step in
labeling the Kerr solution we must impose a Weyl tensor of type D,
and we must obtain the explicit expressions for its scalar
invariants and its underlying geometry. In Section \ref{sec-type-D}
we solve these two questions by making use of the results in
\cite{fms}.

Walker and Penrose \cite{wape} showed that a Killing tensor exists
in the charged Kerr black hole, and Hougston and Sommers \cite{hs1}
proved that, with the exception of the generalized charged
C-metrics, the other charged counterpart of the type D vacuum
solutions also have this property. The type D vacuum solutions with
a Killing tensor are called the vacuum Kerr-NUT metrics. In
\cite{fsEM-sym} we have studied the symmetries and other invariant
properties of the type D vacuum metrics. These results allow us to
give in Section \ref{sec-KN-metrics} an intrinsic labeling of the
Kerr-NUT vacuum solutions. In this section we also offer an
alternative characterization of the vacuum Kerr-NUT metrics more
based on the Weyl scalar invariants.

In Section \ref{sec-faetures-KN} we give the coordinate expression
of the basic Weyl invariants in a Kerr-NUT vacuum solution, and we
also study some restrictions on the Weyl invariants that hold in
these space-times. We need these result in order to obtain new
intrinsic conditions that distinguish the Kerr solution from the
other vacuum Kerr-NUT metrics.

The Kerr characterization theorem is presented in Section
\ref{sec-Kerr}. We offer two alternative intrinsic and explicit
labeling of the Kerr geometry. The second one is more adapted to the
Weyl scalar invariants and it seems suitable to obtain an
algorithmic characterization of the Kerr initial data. We also offer
the invariant expression of the Kerr mass and angular momentum, as
well as, of the `stationary' Killing vector field.

Finally, Section \ref{sec-ending} is devoted to remark the
algorithmic nature of our characterization theorem by presenting a
summary of the results as a flow chart.

In this paper we work on an oriented space-time with a metric tensor
$g$ of signature $\{-,+,+,+\}$. The Ricci and Weyl tensors are
defined as given in \cite{kramer} and denoted, respectively, by
$Ric$ and $W$. For the metric product of two vectors we write $(x,y)
= g(x,y)$, and we put $x^2 = g(x,x)$. Other basic notation used in
this work is summarized in \ref{notation}.

\section{Labeling Petrov-Bel type D space-times} \label{sec-type-D}

The Kerr solution is a Petrov-Bel type D metric. In this section we
give the equations that make explicit this intrinsic condition.
Firstly, in the self-dual complex formalism which is well adapted to
the study of the Weyl eigenvalue problem, and secondly, in real
formalism.

\subsection{Type D metrics in self-dual formalism}

In order to analyze the conditions for a space-time to be a
Petrov-Bel type D solution we now introduce the necessary notation.
A self--dual 2--form is a complex 2--form ${\cal F}$ such that
$*{\cal F}= \textrm{i}{\cal F}$, where $*$ denotes the Hodge dual
operator. We can associate biunivocally with every real 2--form $F$
the self-dual {2--form ${\cal
F}=\frac{1}{\sqrt{2}}(F-\textrm{i}*F)$. Here we refer to a
self--dual 2--form as a {\it SD bivector}. The endowed metric on the
3-dimensional complex space of the SD bivectors is ${\cal
G}=\frac{1}{2}(G-\textrm{i} \; \eta)$, $\eta$ being the metric
volume element of the space-time and $G$ the metric on the space of
2--forms, $G=\frac{1}{2} g \wedge g$.

Every double 2--form, and in particular the Weyl tensor $W$, can be
considered as an endomorphism on the space of the 2--forms. The
restriction of the Weyl tensor on the SD bivectors space is the {\em
self-dual Weyl tensor} and is given by $ {\cal W} = \frac12(W - \ci
*W)$. The algebraic classification of the Weyl tensor $W$ can be obtained
by studying the traceless linear map defined by the self--dual Weyl
tensor ${\cal W}$ on the SD bivectors space \cite{petrov,bel,fms}.
The characteristic equation reads $\, x^{3}-\frac{1}{2} a x
-\frac{1}{3} b =0  \, $, where the complex invariants $a$ and $b$
are given by $a \equiv  \Tr {\cal W}^2$, $b \equiv \Tr {\cal W}^3$.

In a Petrov-Bel type D space-time the self-dual Weyl tensor
satisfies $6b^2=a^3$ and has a minimal polynomial of degree two
\cite{bel}. Then, it has a double eigenvalue $w$ and admits the
canonical expression \cite{fms}:
\begin{equation}  \label{type-D-canonica}
{\cal W} = 3 w \, {\cal U} \otimes {\cal U} + w \, {\cal G} \, ,
\qquad w \equiv - \frac{b}{a}  \, ,
\end{equation}
where ${\cal U}$ is the normalized eigenbivector associated with the
simple eigenvalue $-2w$. The {\em canonical bivector} ${\cal U} =
\frac{1}{\sqrt{2}}(U-\textrm{i}*U)$ determines the two {\em
principal planes} of a type D Weyl tensor. The projector on the
time-like (resp., space-like) principal plane is $v = U^2$ (resp.,
$h = g-v = -(*U)^2 $).

We summarize the results that we need below in two lemmas
\cite{fms}.
\begin{lemma} \label{lemma-type-D}
A space-time is Petrov-Bel type D if, and only if, the self-dual
Weyl tensor satisfies:
\begin{equation}  \label{type-D}
a \not= 0 \, ,  \qquad  {\cal W}^2 - \frac{b}{a}\, {\cal W} -
\frac{a}{3}\, {\cal G} = 0 \, ; \qquad a \equiv  \Tr {\cal W}^2 \, ,
\quad b \equiv \Tr {\cal W}^3 \, .
\end{equation}
\end{lemma}

\begin{lemma} \label{lemma-determ-U}
The eigenvalue $w$ and the SD canonical bivector ${\cal U}$ of a
Petrov-Bel type D Weyl tensor can be obtained as:
\begin{equation} \label{determ-U}
w = - \frac{b}{a} \, , \qquad \quad    {\cal U} = \frac{{\cal
P}(Z)}{\sqrt{-{\cal P}^2(Z,Z)}} \, ; \qquad \quad {\cal P} \equiv
{\cal W} +
\frac{b}{a} \, {\cal G} \, ,
\end{equation}
where $Z$ is an arbitrary bivector.
\end{lemma}

\subsection{Type D metrics in real formalism} \label{subsec-type-D}

Now, we express the statements of the above lemmas in real
formalism. The real Weyl scalar invariants are defined as
\cite{bel}:
\begin{equation} \label{real-invariants}
\hspace{-2cm} A \equiv \frac{1}{2} \Tr W^2 \, , \quad B \equiv
\frac{1}{2} \Tr (W \circ *W) \, , \quad D \equiv \frac{1}{2} \Tr W^3
\, , \quad E \equiv \frac{1}{2} \Tr (W^2 \circ
*W)  \, ,
\end{equation}
and they are related with the complex ones by $a = A - \ci B, \ b =
D - \ci E$.

If we denote $\alpha$ and $\beta$ the real and imaginary parts of
the double Weyl eigenvalue $w$, $w= \alpha + \ci \beta$, we find
that, in terms of the real Weyl scalar invariants, they are given by
\begin{equation} \label{real-eigenvalue}
\alpha= - \frac{A D + B E }{A^2 + B^2} \, , \qquad \quad \beta =
\frac{A E - B D}{A^2 + B^2}  \, .
\end{equation}
Then, from this notation, lemma \ref{lemma-type-D} becomes:
\begin{lemma} \label{lemma-type-D-real}
A space-time is Petrov-Bel type D if, and only if, the Weyl tensor
satisfies:
\begin{equation}  \label{type-D-real}
A^2 + B^2 \not= 0 \, ,  \qquad  W^2 + \alpha W + \beta *W  -
\frac{1}{3} (A G - B \eta) = 0 \, ,
\end{equation}
where $A$, $B$, $\alpha$ and $\beta$ are given in {\rm
(\ref{real-invariants})} and {\rm (\ref{real-eigenvalue})}.
\end{lemma}

From (\ref{type-D-canonica}) we can obtain the canonical expression
for the real Weyl tensor $W$ in terms of the Weyl scalar invariants
and the canonical bivector:
\begin{equation} \label{type-D-canonica-real}
W = 3 \alpha (U \otimes U - *U \otimes *U) +  3 \beta U
\stackrel{\sim}{\otimes} *U + \alpha G + \beta \eta \, .
\end{equation}

On the other hand, the real part $P$ of the SD double-2-form ${\cal
P}$ given in (\ref{determ-U}) projects every bivector $Z$ on the
plane of bivectors generated by $U$ and $*U$. Then a straightforward
calculation leads to:

\begin{lemma} \label{lemma-determ-U-real}
For a Petrov-Bel type D Weyl tensor, the real and imaginary parts of
the Weyl eigenvalue, $\alpha$ and $\beta$, are given by {\rm
(\ref{real-eigenvalue})}, and the canonical bivector $U$ can be
obtained as:
\begin{equation} \label{determ-U-real}
U = \frac{1}{\chi \sqrt{\chi + f}}\left((\chi + f) \, F + \tilde{f}
\, *F\right) \, ; \qquad F \equiv P(Z) \, ,
\end{equation}
where $Z$ is an arbitrary bivector and
\begin{equation}
\hspace{-1cm} P \equiv W - \alpha G - \beta \eta \, , \quad \chi
\equiv \sqrt{f^2 + \tilde{f}^2} \, , \quad f \equiv \tr F^2 \, ,
\quad \tilde{f} \equiv \tr (F \cdot *F) \, . \label{determ-U-real-2}
\end{equation}

\end{lemma}

\section{Intrinsic characterization of the Kerr-NUT vacuum
solutions} \label{sec-KN-metrics}

The Kerr metric is a type D vacuum solution. We have an intrinsic
and explicit expression for the vacuum condition, $Ric(g)=0$. And
now, after our above analysis, we also have an intrinsic and
explicit expression (\ref{type-D}) (or, equivalently,
(\ref{type-D-real})) for the type D condition. Moreover, we have
obtained the explicit expressions (\ref{determ-U}) (or,
equivalently, (\ref{real-eigenvalue}) and
(\ref{determ-U-real}-\ref{determ-U-real-2})) for the Weyl eigenvalue
$w= \alpha + \ci \beta$ and the canonical bivector $U$. Next, we
must use these Weyl concomitants to distinguish the Kerr solution
from the other type D vacuum solutions.

In order to carry out this step we remember some features of the
type D vacuum solutions and of their charged counterpart, the
so-called ${\cal D}$-metrics. Recently \cite{fs-EM-align}, we have
obtained an explicit and intrinsic characterization of the ${\cal
D}$-metrics. Our conditions exclusively involve algebraic
concomitants of the Ricci and Weyl tensors.

An important property of the ${\cal D}$-metrics is that the Weyl
invariant $\xi = w^{-\frac{1}{3}} \ \delta{\cal U}$ is a complex
Killing vector ($\delta$ denotes the exterior co-differential). On
the other hand, Walker and Penrose \cite{wape} showed that a Killing
tensor exists in the charged Kerr black hole, and Hougston and
Sommers \cite{hs1} proved that, with the exception of the
generalized charged C-metrics, the other ${\cal D}$-metrics also
have this property. The ${\cal D}$-metrics with a Killing tensor are
called the Kerr-NUT metrics.

In \cite{fsEM-sym} we have studied the symmetries and other
invariant properties of the ${\cal D}$-metrics. Moreover we have
shown that they can be intrinsically labeled by adding some
differential conditions on the canonical bivector given in
(\ref{determ-U}) to the conditions (\ref{type-D}) that require a
type D Weyl tensor. These restrictions impose that: (i) the
principal null directions determine shear-free geodesic null
congruences, (ii) the principal planes define a Maxwellian
structure, and (iii) the Ricci tensor commutes with ${\cal U}$. An
important fact is that these three conditions identically hold in
the vacuum case \cite{fsD}.

\subsection{Labeling the Kerr-NUT metrics. (i) Using the canonical bivector}
\label{subsec-KN-metrics-1}

In \cite{fsEM-sym} we have also presented some equivalent conditions
that distinguish the charged Kerr-NUT metrics from the other ${\cal
D}$-metrics. The first one is a result by Hougston and Sommers
\cite{hs2}: in a Kerr-NUT metric the complex invariant vector $\xi$
defines a unique Killing direction. The second one states that the
Kerr-NUT metrics are the ${\cal D}$-metrics with a Killing two-form
$\nabla \xi$ aligned with the Weyl geometry, $\, \nabla \xi =
\gamma_1 \, U + \gamma_2 *U$. And the third one imposes the
canonical bivector to satisfy the following first order differential
equation:
\begin{equation} \label{NUT}
\delta U \wedge \delta *U = 0 \, .
\end{equation}

A remark on the second statement: it characterizes the Kerr-NUT
metrics and, in the vacuum case, it leads to the Kerr solution under
asymptotic flatness behavior. This result by Mars \cite{mars1}
\cite{mars2} has recently been used to improve the black hole
uniqueness theorems \cite{Ionescu} and to study the Kerr initial
data \cite{GP-VK-1}.

On the other hand, note that condition (\ref{NUT}) only involves the
canonical bivector, i. e., it is intrinsic. Consequently, we have an
intrinsic and explicit characterization of the Kerr-NUT vacuum
solutions:

\begin{proposition} \label{prop-Kerr-NUT}
The Kerr-NUT vacuum solutions are the space-times with vanishing
Ricci tensor, $Ric(g)=0$, and whose Weyl tensor satisfies {\rm
(\ref{type-D-real})} and {\rm (\ref{NUT})}, where $U$ is the
canonical bivector given in {\rm (\ref{determ-U-real})}.
\end{proposition}

\subsection{Labeling the Kerr-NUT metrics. (ii) Without using the
canonical bivector} \label{subsec-KN-metrics-2}

The intrinsic labeling of the Kerr-NUT metrics given in proposition
\ref{prop-Kerr-NUT} is a direct application of the results in
\cite{fsEM-sym}. The condition (\ref{NUT}) has a simple geometrical
interpretation in terms of the Weyl principal planes. Nevertheless,
its strong dependence on the canonical bivector $U$ could be a
handicap for subsequent applications. Indeed, $U$ is a concomitant
of the Weyl tensor, but in its explicit expression
(\ref{determ-U-real}) appears, necessarily, an arbitrary bivector
$Z$. This fact could make difficult to work with equations, like
(\ref{NUT}), involving $U$.

Here we offer an alternative intrinsic labeling of the Kerr-NUT
vacuum metrics that uses the gradients of the Weyl scalar
invariants. This approach is more suitable to study the Kerr initial
data.

We shall start remembering that, for a vacuum metric, the Bianchi
identity states $\delta {\cal W} =0$. And, for type D metrics, this
condition asserts \cite{fsD} that the Weyl principal structure is
umbilical and imposes on the canonical bivector ${\cal U}$ and the
Weyl eigenvalue $w$ the following differential restriction
\cite{fsD}:
\begin{equation} \label{bianchi}
3 \, {\cal U}(\delta {\cal U} ) = -\, \dif \ln w \, .
\end{equation}

In order to write this equation in real formalism, we consider the
invariants $\rho$ and $\theta$ which are essentially the logarithm
of the modulus and the argument of the eigenvalue. If we put $w =
e^{\rho + \ci \theta}$, then (\ref{bianchi}) becomes:
\begin{equation} \label{bianchi-real}
*U(\delta * U) -U (\delta U) = \frac{2}{3} \, \dif \rho \, ,
\qquad *U(\delta U) + U(\delta *U) = \frac{2}{3} \, \dif \theta \, .
\end{equation}

Note that the argument $\theta$ is not uniquely defined, but its
gradient is, and we can compute $\dif \rho$ and $\dif \theta$ using
the fact that $\dif(\rho + \ci  \theta) = \frac{\dif{w}}{w} =
\frac{\dif{w^2}}{2w^2}$. Then, taking into account that $w^2 =
\frac{a}{6}$, we have the explicit Weyl invariants:
\begin{equation} \label{difroteta}
\frac23 \, \dif \rho = \frac{A \dif A + B \dif B}{3(A^2 + B^2) }
\equiv R \, , \qquad \frac23 \, \dif \theta = \frac{B \dif A - A
\dif B}{3(A^2 + B^2)} \equiv \Theta  \, .
\end{equation}
Then, conditions (\ref{bianchi-real}) can be written as:

\begin{equation} \label{bianchi3}
\delta U = - U(R) - *U(\Theta) \,  , \qquad \delta *U = U(\Theta)
-*U(R)  \, .
\end{equation}

These conditions that relate the first derivatives of both, the Weyl
canonical bivector and the Weyl scalar invariants, allow the
Kerr-NUT invariant condition (\ref{NUT}) to be expressed without an
explicit use of the canonical bivector. Indeed, if we exclude the
B-metrics where $\delta U$ identically vanishes, condition
(\ref{NUT}) states that a scalar $\lambda$ exists such that:
\begin{equation} \label{kerr-NUT-lambda}
\delta * U = \lambda \, \delta U \, .
\end{equation}
Under this assumption, (\ref{bianchi3}) is equivalent to
\begin{equation} \label{bianchi3b}
U(\Theta) = - \lambda U(R) \, , \qquad  *U(R) = \lambda *U(\Theta)
\, .
\end{equation}
These two vectorial equations can be stated as a tensorial one that
is also satisfied by the B-metrics:
\begin{equation} \label{Kerr-NUT-Q-a}
(U \stackrel{\sim}{\otimes} *U)(R, R) + (U \stackrel{\sim}{\otimes}
*U)(\Theta , \Theta)=0  \, ,
\end{equation}
where, for a double 2--form $P$ and two vectors $x,y$, $P(x,y)$
denotes the 2--tensor with componets $\, P(x,y)_{\alpha \beta} =
P_{\alpha \mu \beta \nu}\, x^{\mu} y^{\nu}$. Note that the double
2--form $\, U \stackrel{\sim}{\otimes} *U \,$ can be obtained from
the canonical expression (\ref{type-D-canonica-real}) in terms of
the Weyl tensor and the scalar invariants. Moreover, equation
(\ref{Kerr-NUT-Q-a}) is invariant if we modify $U
\stackrel{\sim}{\otimes}
*U$ by a factor or by the addition of a fully skew-symmetric tensor.
Finally, if we use the expressions (\ref{difroteta}) for $R$ and
$\Theta$ we arrive to the following labeling of the Kerr-NUT vacuum
solutions:
\begin{proposition} \label{prop-Kerr-NUT-Q}
The Kerr-NUT vacuum solutions are the space-times with vanishing
Ricci tensor, $Ric(g)=0$, and whose Weyl tensor satisfies {\rm
(\ref{type-D-real})} and
\begin{equation} \label{Kerr-NUT-Q}
Q(\dif A, \dif A) + Q(\dif B , \dif B )=0  \, , \qquad Q \equiv
\beta \, W + \alpha *W   \, ,
\end{equation}
where $A$, $B$, $\alpha$ and $\beta$ are given in {\rm
(\ref{real-invariants})} and {\rm (\ref{real-eigenvalue})}.
\end{proposition}

It is worth remarking that condition (\ref{NUT}) distinguishes the
Kerr-NUT metrics from the other ${\cal D}$-metrics. Thus, it also
applies in the charged case although here we have only used it for
vacuum (proposition \ref{prop-Kerr-NUT}). Nevertheless, condition
(\ref{Kerr-NUT-Q}) has been obtained by using the Bianchi identities
for the vacuum case. Thus, it only distinguishes the vacuum Kerr-NUT
metrics from the other type D vacuum solutions (proposition
\ref{prop-Kerr-NUT-Q}).

\section{Some remarks on the vacuum Kerr-NUT metrics}
\label{sec-faetures-KN}

Once the Kerr-NUT metrics have been labeled, we need to obtain the
complementary restrictions on the Weyl invariants that allow us to
characterize the Kerr metric. In order to tackle this point in the
next section, here we present same useful features of a family of
vacuum Kerr-NUT metrics which contains the Kerr solution. This
family is defined by the regularity conditions:
\begin{equation} \label{K-N-regular}
Q(\delta U, \delta U) \not= 0 \, , \qquad \quad  v(\delta U , \delta
U) \neq 0
\end{equation}
where $Q$ is given in (\ref{Kerr-NUT-Q}) and $v$ is the projector on
the time-like principal plane. The first regularity condition in
(\ref{K-N-regular}) imposes that the invariant directions defined by
the first derivatives of the Weyl tensor have a non-vanishing
projection on the two principal planes.

On the other hand, the ${\cal D}$-metrics admit, at least, a
commutative two-dimensional group of isometries. From the results in
\cite{fsEM-sym}, it follows that the Killing vectors plane is null
if, and only if, $v(\delta U)$ is a null direction everywhere. Thus,
the second regularity condition in (\ref{K-N-regular}) avoids this
case.

The scalar $\lambda$ introduced in (\ref{kerr-NUT-lambda}) plays an
important role in the intrinsic study of the Kerr-NUT metrics and,
in particular, in the characterization of the Kerr solution. Note
that $\lambda$ is a covariant scalar and, if we exclude the
B-metrics, it can be explicitly obtained as:
\begin{equation} \label{lambda-U}
\lambda = \frac{(s\, ,\delta \!*\!U)}{(s\, , \delta U)} \, ,
\end{equation}
$s$ being an arbitrary vector such that $(s\, , \delta U) \not= 0$ .

Moreover, the scalar invariant $\lambda$ is related with the
derivatives of the algebraic Weyl scalars as a consequence of
restrictions (\ref{bianchi3}) and (\ref{bianchi3b}). Indeed, a
straightforward calculation leads to:
\begin{equation}\label{lambda-K}
K  \lambda^2 - 2 \lambda - K = 0 \, , \qquad  K \equiv  \frac{2 \,
(R, \Theta)}{R^2 - \Theta^2} \, .
\end{equation}

The type D vacuum solutions were first obtained by Kinnersley \cite{kin}. The
explicit integration of their charged counterpart, the ${\cal
D}$-metrics, has been acquired by several authors, and they can be deduced from the Pleba\'nsky and Demia\'nski \cite{ple-dem} line element by means of several
limiting procedures (see \cite{kramer} and references therein and
the recent paper by Griffiths and Podolsk´y \cite{griff-pod} for a
detailed analysis). The vacuum Kerr-NUT metrics satisfying the regularity conditions (\ref{K-N-regular}) admit a coordinate line element that
can be found in \cite{kramer}. Moreover, we can obtain the basic Weyl invariants in
this coordinate system:
\begin{lemma} \label{lemma-canonica-Kerr-NUT}
In the Kerr-NUT vacuum solutions such that $Q(\delta U, \delta U) \not= 0$ and $v(\delta U , \delta U) \neq 0$, the metric line element takes the expression:
\begin{equation}   \label{Kerr-NUT-le}
\displaystyle  \hspace{-2cm} g= - \frac{Y}{y^2 + x^2} (\dif t + x^2
\dif z)^2 + \frac{y^2 + x^2}{Y} \, \dif y^2 + \frac{X}{y^2 + x^2}
(\dif t - y^2 \dif z)^2 + \frac{y^2 + x^2}{X} \, \dif x^2 \, ,
\end{equation}
where
\begin{equation}   \label{Kerr-NUT-le-b}
Y \equiv \epsilon y^2 - 2 \mu y + \gamma \, , \qquad X \equiv -
\epsilon x^2 + 2 \nu x + \gamma > 0 \, .
\end{equation}
Moreover, in this coordinate system the double Weyl eigenvalue $w$
and the principal two-forms $U$ and $*U$ are
\begin{equation} \label{Kerr-NUT-U}
\hspace{-9mm} w =- \frac{\mu + \ci \nu}{(y + \ci x)^3} \, , \quad U=
-(\dif t + x^2 \dif z) \wedge \dif y \, , \quad
*U =(\dif t - y^2 \dif z) \wedge \dif x \, .
\end{equation}
\end{lemma}

From the coordinate expression (\ref{Kerr-NUT-U}) of the basic Weyl
invariants and the expression (\ref{lambda-U}) we have
\begin{equation}  \label{deltaU-1}
\delta U = y\, u \, , \qquad \delta * U  = x \, u \, ; \qquad \quad
\lambda = \frac{x}{y} \, ;
\end{equation}
\begin{equation} \label{deltaU-2}
u\equiv\frac{2}{(x^2 + y^2)^2} \{X(dt - y^2 dz) - Y (dt + x^2 dz)\}
\, .
\end{equation}
Then, we can obtain the coordinate expression of the invariant:
\begin{equation} \label{Omega}
\Omega \equiv \omega (1 + \ci \lambda)^3 = - \frac{1}{y^3} (\mu +
\ci \nu)
\end{equation}

The metric line element of the vacuum Kerr-NUT metrics with null
orbits can also be found in \cite{kramer}. From this, we can obtain
the Weyl complex scalar invariant $\Omega$ and we see that it is a
purely imaginary scalar. Thus, we have:
\begin{lemma} \label{lemma-null-orbits}
A Kerr-NUT vacuum solution has null orbits if, and only if,
$v(\delta U , \delta U) = 0$. \\
Moreover, in this case {\rm (\ref{Omega})} is a purely imaginary
Weyl scalar invariant.
\end{lemma}

\section{The characterization theorem for the Kerr metric}
\label{sec-Kerr}

The Kerr solution is a vacuum Kerr-NUT metric which satisfies the
regularity conditions (\ref{K-N-regular}). Thus, its metric line
element follows from (\ref{Kerr-NUT-le}) for particular values of
the coordinate parameters. More precisely, we have \cite{kramer}:
\begin{lemma} The line element {\rm (\ref{Kerr-NUT-le}-\ref{Kerr-NUT-le-b})}
becomes the Kerr solution if we take $\nu = 0$ and $\epsilon > 0$.
Moreover the Kerr mass and angular momentum are given, respectively,
by:
\begin{equation} \label{mass-angular-1}
{\rm m} = \frac{|\mu|}{\epsilon \sqrt{\epsilon}} \, , \qquad \qquad
\qquad {\rm a} = \frac{\sqrt{\gamma}}{\epsilon}
\end{equation}
\end{lemma}

Then, if we want to label the Kerr metric we must add to the
conditions that characterize the vacuum Kerr-NUT metrics
(propositions \ref{prop-Kerr-NUT} or \ref{prop-Kerr-NUT-Q}): (i)
regularity conditions (\ref{K-N-regular}), (ii) intrinsic and
explicit expressions of the coordinate restrictions $\nu = 0$ and
$\epsilon > 0$.

\subsection{Labeling the Kerr solution. (i) Using the
canonical bivector} \label{subsec-Kerr-1}

From the coordinate expression (\ref{Omega}) for the complex scalar
$\Omega$, we obtain that this scalar is real if, and only if,
$\nu=0$. Then, if we obtain the imaginary part of the invariant
expression of $\Omega$, we have:
\begin{equation} \label{escalar-nu}
\nu =0 \qquad \longleftrightarrow \qquad (1 - 3 \lambda^2)\, \beta +
\lambda (3 - \lambda^2) \, \alpha = 0
\end{equation}
Moreover, this invariant condition avoids the case of null orbits as
a consequence of lemma \ref{lemma-null-orbits}.

In order to write the restriction $\epsilon > 0$ intrinsically, we
define the explicit Weyl scalar invariant:
\begin{equation} \label{sigma-1}
\sigma \equiv \frac{2 \,\alpha}{3 \lambda^2 -1} - \frac14 (\delta U)^2
\, .
\end{equation}
From (\ref{Kerr-NUT-U}-\ref{deltaU-1}-\ref{deltaU-2}) we can compute
the coordinate expression of $\sigma$ and we obtain:
\begin{equation}
\sigma = \frac{y^2}{(x^2 + y^2)^2} \, \epsilon
\end{equation}
Thus, $\epsilon > 0$ if, and only if, $\sigma > 0$. On the other
hand, note that we can consider $s= \delta U$ in expression
(\ref{lambda-U}) because $\delta U$ is a non null vector almost
everywhere as follows from the regularity conditions
(\ref{K-N-regular}).

Then, from these considerations and proposition \ref{prop-Kerr-NUT},
we arrive to the following intrinsic and explicit characterization
of the Kerr geometry:
\begin{theorem} \label{theor-kerr-1}
A gravitational field $g$ is the Kerr solution if, and only if, its
Ricci tensor vanishes, $Ric(g)=0$, and its Weyl tensor satisfies the
algebraic conditions:
\begin{equation}  \label{type-D-real-theo1}
A^2 + B^2 \not= 0 \, ,  \qquad  W^2 + \alpha W + \beta
*W - \frac{1}{3} (A G - B \eta) = 0 \, ,
\end{equation}
and the first order differential restrictions:
\begin{equation} \label{theo1-differential}
\begin{array}{ll}
\delta U \wedge \delta *U = 0 \, , & \qquad Q(\delta U, \delta U) \not= 0 \, ,\\[2mm]
(1 - 3 \lambda^2)\, \beta + \lambda (3 - \lambda^2) \, \alpha = 0 \, ,  &
\qquad \sigma > 0 \, ,
\end{array}
\end{equation}
$Q = Q(g)$, $\lambda = \lambda(g)$ and $\, \sigma=\sigma(g)$ being the Weyl
concomitants:
\begin{equation} \label{lambda-sigma}
Q \equiv \beta \, W + \alpha *W   \, , \quad \lambda \equiv \frac{(\delta U,\delta *U)}{(\delta U)^2}   \, ,
\quad \sigma \equiv \frac{2 \, \alpha}{3 \lambda^2 -1} - \frac14
(\delta U)^2 \, ,
\end{equation}
where $\, A=A(g),$ $\, B=B(g),$ $\, D=D(g)$ and $\, E=E(g),$ $\,
\alpha = \alpha(g)$ and $\, \ \beta = \beta(g),$ and $\, U=U(g)$ are
the explicit Weyl concomitants given in {\rm
(\ref{real-invariants})}, {\rm (\ref{real-eigenvalue})} and {\rm
(\ref{determ-U-real}-\ref{determ-U-real-2})}, respectively, and $2G
\equiv g \wedge g$.
\end{theorem}

On the other hand, we can compute the Killing (invariant) vector
$\xi = w^{-\frac{1}{3}} \ \delta{\cal U}$ for the Kerr metric. In
this case it is real and coincides with the stationary Killing
vector in the outside region of the Kerr geometry.

Moreover, from the coordinate expressions of the Weyl invariants
given in section \ref{sec-faetures-KN} we can obtain the invariant
expression of the coordinate parameters, and using
(\ref{mass-angular-1}), the invariant expression of the Kerr mass
and angular momentum. Then, we reach:
\begin{proposition}
In terms of the Weyl invariants defined in theorem {\rm
\ref{theor-kerr-1}}, the Kerr mass ${\rm m}$ and angular momentum
${\rm a}$, and the stationary Killing vector $\xi$ are given,
respectively, by:
\begin{equation} \label{m-a-xi}
\hspace{0mm} {\rm m} = \frac{|\alpha|}{\sigma \sqrt{\sigma}|3
\lambda^2 -1|} \, , \quad  {\rm a} = \frac{1}{2 \sigma \sqrt{1 +
\lambda^2}}\left[h(\delta U , \delta U) + \frac{4 \sigma
\lambda^2}{1 + \lambda^2} \right]^{1/2} ;
\end{equation}
\begin{equation} \label{m-a-xi-b}
\xi= \left( \frac{1-3 \lambda^2}{\alpha} \right)^\frac{1}{3}
\frac{1}{\sqrt{2}}\, \, \delta U \, .
\end{equation}
\end{proposition}

\subsection{Labeling the Kerr solution. (ii) Without using the
canonical bivector} \label{subsec-Kerr-2}

Now we offer an alternative characterization of the Kerr geometry
which could be useful in subsequent applications because it avoids
the explicit calculation of $U$.

At first, the two first conditions in (\ref{theo1-differential}) can
be replaced by the equivalent ones $Q(\dif A , \dif A )= - Q(\dif B,
\dif B)\not=0$. This equality ensures that $\delta U$ and $\delta
*U$ are proportional vectors. We know that the proportionality
factor $\lambda$ fulfills (\ref{lambda-K}) and that, for the Kerr
metric, it must also satisfy equation (\ref{escalar-nu}). If we use
(\ref{lambda-K}) to obtain $\lambda^2 $ and $\lambda^3$ in terms of
$\lambda$ and $K$, and remove them in (\ref{escalar-nu}), we obtain
$\lambda$ as a function of $K$ and $T\equiv \frac{\beta}{\alpha}$
that must be satisfied if $\nu =0$ as happens in the Kerr metric.
But this is not a sufficient condition because $-\frac{1}{\lambda}$
also satisfies (\ref{lambda-K}). It can be shown that this case
leads to  $\mu =0$ in (\ref{Kerr-NUT-le}). Consequently, we need
another condition to distinguish which of the two solution of
(\ref{lambda-K}), $\lambda$ and $-\frac{1}{\lambda}$, satisfies
(\ref{escalar-nu}). To that end, let us define:
\begin{equation} \label{Xi}
\Xi = \frac{1}{(1 - \lambda^2) \sqrt{A^2 + B^2}}  \Big[ \Pi(R,R) -
\Pi(\Theta, \Theta) + Q(R , \Theta) + Q(\Theta, R) \Big]
\end{equation}
where $\Pi \equiv \alpha \, W - \beta *W  - (\alpha^2 + \beta^2) G$. A straightforward calculation shows that $\Xi$ is a non-negative (respectively, non-positive ) quadratic form when $\nu=0$ (respectively, $\mu =0$). Then, the Kerr solution follows if the semi-definite quadratic
form $\Xi$ is non-negative. In order to impose this condition, it is enough to take
the trace with an arbitrary elliptical metric associated to $g$.

On the other hand, from (\ref{bianchi3}) and (\ref{bianchi3b}) we
can determine $(\delta U)^2$ in terms of
$\lambda$, $R$ and $\Theta$, and we can replace them in the
expressions (\ref{sigma-1}) of the scalar
$\sigma$. Then, we obtain the following labeling of the Kerr geometry:

\begin{theorem} \label{theor-kerr-2}
A gravitational field $g$ is the Kerr solution if, and only if, its
Ricci tensor vanishes, $Ric(g)=0$, and its Weyl tensor satisfies the
algebraic conditions:
\begin{equation}  \label{type-D-real-theo2}
A^2 + B^2 \not= 0 \, ,  \qquad  W^2 + \alpha W + \beta
*W - \frac{1}{3} (A G - B \eta) = 0 \, ,
\end{equation}
and the first order differential restrictions:
\begin{equation} \label{theo2-differential}
\begin{array}{ll}
Q(\dif A, \dif A) = - Q(\dif B, \dif B) \neq 0 \, , & \qquad 2\,
\Xi(x,x) +
\tr \Xi > 0 \, ,\\[2mm]
(1 - 3 \lambda^2)\, \beta + \lambda (3 - \lambda^2) \, \alpha = 0 \, ,  &
\qquad \sigma > 0 \, ,
\end{array}
\end{equation}
$x$ being an arbitrary timelike unitary vector, and $\Xi = \Xi(g)$, $\, Q = Q(g)$, $\, \sigma=\sigma(g)$ and $\, \lambda = \lambda(g)$
being the Weyl concomitants:
\begin{equation} \label{Xi-teo}
\hspace{-2cm} \Xi \equiv \frac{1}{(1 - \lambda^2) \sqrt{A^2 + B^2}}  \Big[ \Pi(R,R) -
\Pi(\Theta, \Theta) + Q(R , \Theta) + Q(\Theta, R) \Big] \, ,
\end{equation}
\begin{equation} \label{Q-sigma}
\hspace{-2cm} Q \equiv \beta \, W + \alpha *W   \, , \qquad \qquad \Pi \equiv \alpha \, W - \beta *W  - (\alpha^2 + \beta^2) G \, ;
\end{equation}
\begin{equation} \label{lambda-T-K}
\hspace{-2cm} \sigma \equiv \frac{2 \,
\alpha}{3 \lambda^2 -1} - \frac{R^2- \Theta^2}{4 (\lambda^2 - 1)}\, ; \quad \lambda \equiv \frac{ K(K T+ 1)}{  K^2 - 3 K \, T- 2}\, , \quad T
\equiv \frac{\beta}{\alpha}   \, , \quad K \equiv  \frac{2 \, (R,
\Theta)}{R^2 - \Theta^2} \, ,
\end{equation}
where $\, A=A(g),$ $\, B=B(g),$ $\, D=D(g)$ and $\, E=E(g),$ $\,
\alpha = \alpha(g)$ and $\, \ \beta = \beta(g),$ and $\, R=R(g)$ and
$\, \Theta=\Theta(g)$ are the explicit Weyl concomitants given in
{\rm (\ref{real-invariants})}, {\rm (\ref{real-eigenvalue})} and
{\rm (\ref{difroteta})}, respectively, and $2G \equiv g \wedge g$.
\end{theorem}
Moreover, we can obtain the invariant vector $\delta U$ that gives
the Killing vector $\xi$ in (\ref{m-a-xi}) without making use of the
expression of the canonical bivector $U$. Indeed, from the
expressions (\ref{type-D-canonica-real}), (\ref{bianchi3}) and (\ref{bianchi3b}), we obtain that, under the conditions of theorem \ref{theor-kerr-2}, the quadratic form (\ref{Xi-teo}) satisfies $2 \, \Xi = \delta U \otimes
\delta U$. On the other hand, from (\ref{bianchi3}) and (\ref{bianchi3b}) we
can determine $h(\delta U, \delta U)$ in terms of
$\lambda$, $R$ and $\Theta$, and we can replace them in the
expressions (\ref{m-a-xi}) of the angular momentum ${\rm a}$. Then, we can state:

\begin{proposition}
In terms of the Weyl invariants defined in theorem {\rm
\ref{theor-kerr-2}}, the Kerr mass ${\rm m}$ and angular momentum
${\rm a}$, and the stationary Killing vector $\xi$ are given,
respectively, by:
\begin{equation} \label{m-a-xi-2a}
%\hspace{-22mm}
{\rm m} = \frac{|\alpha|}{\sigma \sqrt{\sigma}|3 \lambda^2 -1|} \, ,
\qquad  {\rm a} = \frac{1}{2 \sigma (1 + \lambda^2)}\left[\lambda
(R,\Theta) + \Theta^2 + 4 \sigma \lambda^2 \right]^{1/2}  \, ;
\end{equation}
\begin{equation} \label{m-a-xi-2b}
\xi= \left( \frac{1-3 \lambda^2}{\alpha} \right)^\frac{1}{3}
\frac{\Xi(x)}{\sqrt{\Xi(x,x)}} \, ,
\end{equation}
where $x$ is an arbitrary vector such that $\Xi(x) \not=0$.
\end{proposition}

\section{Summary and ending comments}
\label{sec-ending}

In this work we present the intrinsic and explicit labeling of the
Kerr solution in two different versions (theorems \ref{theor-kerr-1}
and \ref{theor-kerr-2}). These results extend the intrinsic labeling
of the Schwarzschild geometry given in \cite{fsS} to the rotating
case. The algorithmic nature of this last result has enabled an
algorithmic characterization of the Schwarzschild initial data to be
given \cite{GP-VK-2}. Similarly, our Kerr characterization theorem
must be a basic tool in giving a fully algorithmic characterization
of the Kerr initial data which could improve the results presented
in \cite{GP-VK-1}.

The algorithmic nature of our results is a direct consequence of
their intrinsic and explicit presentation. We make this algorithmic
nature more evident by summarizing the Kerr characterization in a
flow chart (see below). The diagram shows the role played by every
condition in the theorem. Note that the algebraic conditions label
the type D vacuum solutions. The first differential equation in
(\ref{theo1-differential}) distinguishes the Kerr-NUT metrics from
the C-metrics. Finally, the other the three differential conditions
in (\ref{theo1-differential}) determine the Kerr geometry.

It is worth remarking that Theorem \ref{theor-kerr-2} labels the
Schwarzschild geometry as a limit case. Indeed, we can easily prove
that, if we remove in theorem 2 the regularity requirement $Q(\dif
B, \dif B) \not= 0$, the three conditions $\beta=0$, $\lambda=0$,
${\rm a} = 0$, are equivalent restrictions on the Weyl tensor. Then,
the other conditions in theorem \ref{theor-kerr-2} become the
conditions of the Schwarzschild characterization theorem given in
\cite{fsS}. Moreover, the invariant expressions (\ref{m-a-xi-2a})
for ${\rm m}$ and (\ref{m-a-xi-2b}) for $\xi$ become those given in
\cite{fsS} for the Schwarzschild mass and the static Killing vector.

\vspace{1.2cm}

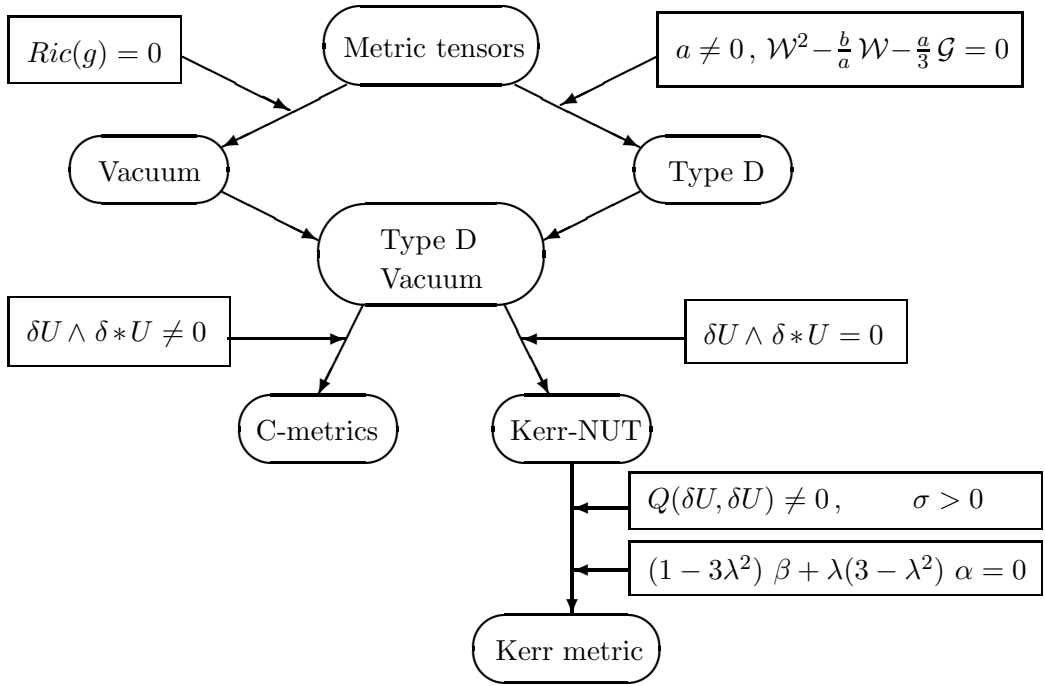
\setlength{\unitlength}{0.75cm} { \hspace{0.3cm} \noindent
\begin{picture}(0,20)
\thicklines
\small

 \put(-0.5,18){\fboxrule
0.5pt \fbox{
  \parbox{55pt}{\vspace{2mm}
$Ric(g) = 0$\vspace{0.7mm}}}}

\put(11,18){\fboxrule 0.5pt \fbox{
    \parbox{128pt}{\vspace*{1.4mm}
$  a \not= 0 \, ,   \,    {\cal W}^2 - \frac{b}{a}\, {\cal W} -
\frac{a}{3}\, {\cal G} = 0 \, $\vspace*{1.4mm}}}}

 \put(7,18.2){\oval(3.8,1.4 )}
\put(5.45,18){\mbox{Metric tensors}}

 \put(2,16){\oval(2.8,1.2 )} \put(12,16){\oval(2.8,1.2)}

\put(1.1,15.8){\mbox{Vacuum}} \put(11.2,15.8){\mbox{Type
 D}}

\put(7,14.5){\oval(4,1.8 )}

 \put(6.1,14.6){\mbox{Type D}}
\put(6.1,13.9){\mbox{Vacuum}}

\put(11.5,13){\fbox{
    \parbox{70pt}{\vspace{2mm}
$\delta U \wedge \, \delta *U = 0\, $\vspace{1.5mm}} }}

\put(-0.5,13){\fbox{
    \parbox{70pt}{\vspace{2mm}
$\delta U \wedge \, \delta *U \neq 0\, $\vspace{1.5mm}} }}

\put(5 ,11.4){\oval(2.8,1.2 )} \put(9.5,11.4){\oval(2.8,1.2)}
\put(3.9,11.2){\mbox{C-metrics}} \put(8.4,11.2){\mbox{Kerr-NUT }}

\put(9.4,7.5){\oval(3.3,1.2)} \put(8.14,7.3){\mbox{Kerr metric}}

\put(9.5,10.8){\vector(0,-1){2.7 }}

 \put(10.5,10){\fboxrule 0.5pt \fbox{
    \parbox{146pt}{\vspace{0.7mm}
$Q(\delta U, \delta U) \not= 0 \, , \qquad  \sigma > 0$\vspace{0.7mm}}}}

\put(2.6,18){\vector(2,-1){1.9 }} \put(11,18){\vector(-2,-1){1.7}}
\put(5.5,17.5){\vector(-2,-1){2.2 }}
\put(8.5,17.5){\vector(2,-1){2.2 }}

 \put(3.3,15.6){\vector(2,-1){1.7}}

 \put(10.7,15.6){\vector(-2,-1){1.7}}

\put(3.4,13){\vector(1,0){2.1 }} \put(11.5,13){\vector(-1,0){2.9}}

\put(5.8,13.6){\vector(-1,-2){.8 }}
\put(8.3,13.6){\vector(1,-2){.8}}

\put(10.5,8.9){\vector(-1,0){1 }} \put(10.5,10){\vector(-1,0){1 }}

\put(10.5,8.8){\fboxrule 0.5pt \fbox{
    \parbox{146pt}{\vspace{0.7mm}
$(1 - 3 \lambda^2) \ \beta +\lambda (3 - \lambda^2)\ \alpha =0 $}} }
\end{picture}
}

\vspace{-4.4cm}

\normalsize

\ack This work has been supported by the Spanish Ministerio de
Educaci\'on y Ciencia, MEC-FEDER project FIS2006-06062.

\appendix

\section{Notation}
\label{notation}

\begin{enumerate}
\item
{\bf Products and other formulas involving 2-tensors $A$ and $B$}:
\begin{enumerate}
\item
Composition as endomorphisms: $A \cdot B$,
\begin{equation}
(A \cdot B)^{\alpha}_{\ \beta} = A^{\alpha}_{\ \mu} B^{\mu}_{\
\beta}
\end{equation}
\item
Square and trace as an endomorphism:
\begin{equation}
A^2 = A \cdot A \, , \qquad \tr A = A^{\alpha}_{\ \alpha}.
\end{equation}
\item
Action on a vector $x$, as an endomorphism $A(x)$, and as a
quadratic form $ A(x,x)$:
\begin{equation}
A(x)^{\alpha} = A^{\alpha}_{\ \beta} x^{\beta}\, , \qquad A(x,x) =
A_{\alpha \beta} x^{\alpha} x^{\beta} \, .
\end{equation}
\item
Exterior product as double 1-forms:  $(A \wedge B)$,
\begin{equation}
(A \wedge B)_{\alpha \beta \mu \nu} = A_{\alpha \mu} B_{\beta \nu} +
A_{\beta \nu} B_{\alpha \mu} - A_{\alpha \nu} B_{\beta \mu} -
A_{\beta \mu} B_{\alpha \nu} \, .
\end{equation}
\end{enumerate}
\item
{\bf Products and other formulas involving double 2-forms $P$ and
$Q$}:
\begin{enumerate}
\item
Composition as endomorphisms of the bivectors space: $P \circ Q$,
\begin{equation}
(P \circ Q)^{\alpha \beta}_{\ \ \rho \sigma} = \frac12 P^{\alpha
\beta}_{\ \ \mu \nu} Q^{\mu \nu}_{\ \ \rho \sigma}
\end{equation}
\item
Square and trace as an endomorphism:
\begin{equation}
P^2 = P \circ P \, , \qquad \Tr P = \frac12 P^{\alpha \beta}_{\ \
\alpha \beta}.
\end{equation}
\item
Action on a bivector $X$, as an endomorphism $P(X)$, and as a
quadratic form $P(X,X)$,
\begin{equation}
P(X)^{\alpha \beta} = \frac12 P^{\alpha \beta}_{\ \ \mu \nu} X^{\mu
\nu} \, , \qquad P(X,X) = \frac14 P_{\alpha \beta \mu \nu} X^{\alpha
\beta} X^{\mu \nu}.
\end{equation}
\item
The Hodge dual operator is defined as the action of the metric
volume element $\eta$ on a bivector $F$ and a double 2-form $W$:
\begin{equation}
*F = \eta(F)  \, , \qquad  *W = \eta \circ W \, .
\end{equation}
\end{enumerate}
\end{enumerate}


\section*{References}
\begin{thebibliography}{999}

\bibitem{GP-VK-1} Garc\'{\i}a-Parrado G\'omez-Lobo A and  Valiente Kroon J A 2008
{\it Class. Quantum Grav.} {\bf 25} 205018

\bibitem{mars1} Mars M 1999  {\it Class. Quantum Grav.} {\bf 16} 2507

\bibitem{mars2} Mars M 2000  {\it Class. Quantum Grav.} {\bf 17} 3353

\bibitem{GP-VK-2}  Garc\'{\i}a-Parrado G\'omez-Lobo A and  Valiente Kroon J
A 2007 {\it Phys. Rev. D} {\bf 75} 024027

\bibitem{fsS} Ferrando J J and S\'aez J A 1998 {\it Class. Quantum Grav.}
{\bf 15} 1323

\bibitem{simon} Simon W 1984 {\it Gen. Rel. Grav.} {\bf 16} 465

\bibitem{Ionescu} Ionescu A D and Klainerman S 2007 (arXiv:0711.0040 [gr-qc])

\bibitem{Riemann} Riemann B 1867 {\it Abhandlungen der K\"oniglichen
Gesellschaft der Wissenschaften zu G\"ottingen} {\bf 13} 272

\bibitem{Bianchi} Bianchi L 1902 {\it Lezioni di geometria
differenziale} vol 2 (Spoerri, Pisa)

\bibitem{Cotton} Cotton \'E 1899 {\it Annales de la facult\'e des
sciences de Toulouse} 2e s\'erie {\bf 1} 385

\bibitem{Weyl} Weyl H 1918 {\it Mathematische Zeitschrift} {\bf 2}
384

\bibitem{Schouten} Schouten J A 1921 {\it Mathematische Zeitschrift} {\bf 11}
58

\bibitem{cartan} Cartan E 1946 {\it Le\c{c}ons sur le G\'{e}om\'{e}trie
des Espaces de Riemann} (Gauthier-Vilar, Paris)

\bibitem{brans} Brans C H 1965 {\it J. Math. Phys.} {\bf 6} 94

\bibitem{karlhede} Karlhede A 1980 {\it Gen. Rel. Grav.} {\bf 12} 693

\bibitem{churchill} Churchill R V 1932 {\it Trans. Am. Math. Soc}
{\bf 34} 784

\bibitem{plebanski} Pleba\'nski J 1964 {\it Acta Phys. Polon.} {\bf
B 11} 579

\bibitem{bcm} Bona C, Coll B and Morales J A 1992 {\it J. Math. Phys.} {\bf 33} 670

\bibitem{petrov} Petrov A Z 1954 {\it Sci. Not. Kazan Univ.} {\bf 114} 55.
This article has been reprinted in 2000 {\it Gen. Rel. Grav.} {\bf
32} 1665

\bibitem{bel} Bel L 1962 {\it Cah. de Phys.} {\bf 16} 59. This article
has been reprinted in 2000 {\it Gen. Rel. Grav.} {\bf 32} 2047
(2000)

\bibitem{fms} Ferrando J J, Morales J A  and S\'aez J A 2001 {\it Class.
Quantum Grav.} {\bf 18} 4969

\bibitem{wape} Walker M and Penrose R,  1970 {\it Commun. Math. Phys.}
{\bf 18} 265

\bibitem{hs1} Hougshton L P and Sommers P 1973 {\it Commun. Math. Phys.}
{\bf 32} 147

\bibitem{fsEM-sym} Ferrando J J and  S\'aez J A 2007 {\it J. Math. Phys.}
{\bf 48} 102504

\bibitem{kramer} Stephani E, Kramer H, McCallum M A H, Hoenselaers C and Hertl E 2003
{\it Exact Solutions of Einstein's Field Equations} (Cambridge
University Press, Cambridge)

\bibitem{fs-EM-align}  Ferrando J J and  S\'aez J A 2007 {\it Gen. Relativ.
Gravit.}  {\bf 39} 2039

\bibitem{fsD}  Ferrando J J and  S\'aez J A 2004 {\it J. Math. Phys.} {\bf 45} 652

\bibitem{hs2} Hougshton L P and Sommers P 1973 {\it Commun. Math. Phys.}
{\bf 33} 129

\bibitem{kin} Kinnersley  W 1969 {\it J. Math. Phys.} {\bf 10} 1195

\bibitem{ple-dem} Pleba\'nski J F and Demia\'nski M 1976 {\it Ann. Phys.
(NY)} {\bf 98} 98

\bibitem{griff-pod} Griffiths J B and Podolsk\'y J 2006 {\it Int. J. Mod. Phys. D} {\bf 15},
335


\end{thebibliography}
\end{document}